\documentclass[prd,aps,twocolumn,preprintnumbers, showpacs, nofootinbib,superscriptaddress,notitlepage]{revtex4-1}
\usepackage{amssymb, amsthm}
\usepackage{amsmath}    
\usepackage{mathrsfs}   
\usepackage{graphicx}   
\usepackage{color}      
\usepackage{footnote}   

\usepackage{slashed}    
\usepackage{subfigure}  
\usepackage{hyperref}   

\usepackage{rotating}   
\usepackage{multirow}   

\newcommand{\beq}{\begin{eqnarray}}
\newcommand{\eeq}{\end{eqnarray}}

\begin{document}

\preprint{MSUHEP-18-003,MIT-CTP/4991}

\title{Lattice Calculation of Parton Distribution Function from LaMET at Physical Pion Mass with Large Nucleon Momentum}

\collaboration{\bf{Lattice Parton Physics Project ($\rm {\bf LP^3}$) Collaboration}} 

\author{Jiunn-Wei Chen}
\affiliation{Department of Physics, Center for Theoretical Physics, and Leung Center for Cosmology and Particle Astrophysics, National Taiwan University, Taipei, Taiwan 106}

\author{Luchang Jin}
\affiliation{Physics Department, University of Connecticut,
Storrs, Connecticut 06269-3046, USA}
\affiliation{RIKEN BNL Research Center, Brookhaven National Laboratory,
Upton, NY 11973, USA}

\author{Huey-Wen Lin}
\affiliation{Department of Physics and Astronomy, Michigan State University, East Lansing, MI 48824}
\affiliation{Department of Computational Mathematics, Michigan State University, East Lansing, MI 48824}

\author{Yu-Sheng Liu}
\affiliation{Tsung-Dao Lee Institute, Shanghai 200240, China}

\author{Yi-Bo Yang}
\affiliation{Department of Physics and Astronomy, Michigan State University, East Lansing, MI 48824}

\author{Jian-Hui Zhang}
\affiliation{Institut f\"ur Theoretische Physik, Universit\"at Regensburg, D-93040 Regensburg, Germany}

\author{Yong Zhao}
\affiliation{Center for Theoretical Physics, Massachusetts Institute of Technology, Cambridge, MA 02139, USA}

\date{March 11, 2018}

\begin{abstract}
We present a lattice-QCD calculation of the unpolarized isovector parton distribution function (PDF) using ensembles at the physical pion mass with large proton boost momenta $P_z \in \{2.2,2.6,3.0\}$~GeV within the framework of large-momentum effective theory (LaMET).
In contrast to our previous physical-pion PDF result, we increase the statistics significantly, double the boost momentum, increase the investment in excited-state contamination systematics, and switch to $\gamma_t$ operator to avoid mixing with scalar matrix elements. We use four source-sink separations in our analysis to control the systematics associated with excited-state contamination. 
The one-loop LaMET matching corresponding to the new operator is calculated and applied to our lattice data. We detail the systematics that affect PDF calculations, providing guidelines to improve the precision of future lattice PDF calculations.  We find our final parton distribution to be in reasonable agreement with the PDF provided by the latest phenomenological analysis.
\end{abstract}
\maketitle

{\bf Introduction:} Parton distribution functions (PDFs) are important quantities describing the
probability densities of quarks and gluons within hadrons. They are not only fundamental properties of 
quantum chromodynamics (QCD), 
but also are key inputs to predict cross sections in high-energy scattering experiments~\cite{Alekhin:2017kpj}. Calculating the $x$-dependence of PDFs from first principles has long been a holy grail in nuclear and high-energy physics. In modern parton physics, the PDFs are defined from the lightcone correlations of quarks and gluons in the hadron, so they involve strong infrared (IR) dynamics and can only be solved by nonperturbative methods such as lattice QCD. However, the direct calculation of PDFs on a Euclidean lattice has been extremely difficult because the real-time dependence of the lightcone makes it infeasible to extract them from lattice simulations with imaginary time. Early studies based on the operator product expansion (OPE) can only access the lower moments of the PDF~\cite{Martinelli:1987zd,Martinelli:1988xs,Detmold:2001dv,Dolgov:2002zm} from lattice QCD.  A similar situation also occurs in lattice calculations of other parton observables, such as the distribution amplitudes (DAs) and generalized parton distributions (GPDs).

In recent years, it was proposed~\cite{Ji:2013fga,Ji:2013dva,Hatta:2013gta,Ji:2014gla,Ji:2014lra} that the PDFs and other parton observables can be directly extracted from the lattice by calculating the matrix elements of certain static operators in a boosted hadron state. For the unpolarized quark PDF, the static operator is $O_{\Gamma}(z)=\bar{\psi}(z)\Gamma U(z, 0) \psi (0)$ where the space-like Wilson line $U(z, 0)= P\exp\left(-ig\int_0^z dz' A_z(z')\right)$, and $\Gamma=\gamma_z$ or $\gamma_t$ so that under the infinite Lorentz boost along the $z$-axis $O_\Gamma$ approaches the lightcone correlation operator which defines the PDF. The hadron matrix element of $O_{\Gamma}(z)$ 
can be directly obtained from lattice QCD and its Fourier transformation
is known as the quasi-PDF:
\begin{equation}\label{eq:quasipdf}
\tilde{q}_\Gamma(x, P_z, \tilde{\mu})
= \int_{-\infty}^\infty \frac{dz}{2\pi}\ e^{ixP_zz} \big\langle P \big| O_{\Gamma}(z) \big|P\big\rangle\,.
\end{equation} 
The quasi-PDF is related to the lightcone PDF through a factorization theorem, where the former can be factorized into a perturbative matching coefficient and the latter, up to power corrections suppressed by the nucleon momentum. This factorization theorem is founded in the large-momentum effective theory (LaMET)~\cite{Ji:2013fga,Ji:2013dva,Hatta:2013gta,Ji:2014gla,Ji:2014lra}, where the matching coefficient can be calculated exactly in perturbation theory.

Since the proposal of LaMET, much progress has been achieved in the theory side, including the matching coefficient connecting the quasi-PDFs to the PDFs at one-loop order~\cite{Xiong:2013bka,Ji:2015jwa,Ji:2015qla,Xiong:2015nua,Ji:2014hxa,Ji:2018hvs,Stewart:2017tvs,Constantinou:2017sej,Green:2017xeu,
Izubuchi:2018srq,Wang:2017qyg,Wang:2017eel}, the nucleon-mass correction~\cite{Chen:2016utp}, and the renormalization properties of the quasi-PDF~\cite{Ji:2015jwa,Ishikawa:2016znu,Chen:2016fxx,Xiong:2017jtn,Constantinou:2017sej,Ji:2017oey,Ishikawa:2017faj,Green:2017xeu,Chen:2017mzz}. There has also been progress in lattice simulation, progressing from the isovector quark PDF of the nucleon~\cite{Lin:2014zya,Alexandrou:2015rja,Chen:2016utp,Alexandrou:2016jqi,Lin:2017ani} to the meson DAs~\cite{Zhang:2017bzy,Chen:2017gck} and nonperturbative renormalization (NPR) in the regularization-independent momentum subtraction (RI/MOM) scheme~\cite{Chen:2017mzz,Alexandrou:2017huk}. Certain technical issues regarding the nonperturbative renormalization were raised and addressed in Refs.~\cite{Constantinou:2017sej,Alexandrou:2017huk,Green:2017xeu,Chen:2017mzz,Chen:2017mie,Lin:2017ani,Chen:2017lnm}. 
In parallel, there have also been other proposals to calculate the PDFs in lattice QCD~\cite{Ma:2014jla,Ma:2017pxb,Radyushkin:2017cyf,Orginos:2017kos,Liu:1993cv,Liang:2017mye,Detmold:2005gg,Braun:2007wv,Bali:2017gfr,Chambers:2017dov} which are subject to their own systematics, but they can be complementary to each other as well as the LaMET approach.

So far, all the lattice calculations of the unpolarized quasi-PDF except Ref.~\cite{Alexandrou:2018pbm} have been done for $O_{\gamma_z}$, which mixes with $O_{\cal I}$ at $O(a^0)$. The operator mixing introduces an additional systematic uncertainty in the nonperturbative renormalization that is not negligible~\cite{Alexandrou:2017huk,Chen:2017mzz,Green:2017xeu,Lin:2017ani}, thus limiting the precision of the extracted PDF. Fortunately, according to Refs.~\cite{Constantinou:2017sej,Chen:2017mzz,Chen:2017mie}, the $O_{\gamma_t}$ case is free from operator mixing at $O(a^0)$, although the mixing still exists at $O(a)$~\cite{Chen:2017mie}. Therefore, it is highly desirable to start from the quasi-PDF with $O_{\gamma_t}$ to improve the systematic uncertainty from operator mixing in the renormalization procedure.

In this work, we describe a state-of-the-art lattice calculation of the unpolarized isovector quark distribution at physical pion mass. We calculate the quasi-PDF with nucleon momenta as large as 3.0~GeV and use 4 source-sink separations to remove excited-state contamination. The nucleon matrix elements are renormalized using RI/MOM scheme; a new matching calculation is done with the new operator $O_{\gamma_t}$ to connect the RI/MOM quasi-PDF to the $\overline{\text{MS}}$ renormalized lightcone PDF; the nucleon mass correction is included. Our result shows a significant improvement compared to previous lattice studies, in particular in the moderate to large-$x$ region. It also signals a promising trend that the precision of lattice calculations is approaching the precision of phenomenological studies with increasing computing resources.

{\bf Nonperturbative Renormalization and Matching:} 
A nonperturbative renormalization on the lattice is required to obtain the continuum limit of the quasi-PDF matrix element, which is subject to linear UV divergences. In this work, we follow the RI/MOM scheme elaborated in Refs.~\cite{Stewart:2017tvs,Chen:2017mzz}, and match the result to the $\overline{\text{MS}}$ PDF with the one-loop matching coefficient calculated with the method developed in Ref.~\cite{Stewart:2017tvs}.

First, the RI/MOM renormalization constant $\tilde{Z}$ is calculated nonperturbatively from the lattice by imposing the following momentum subtraction condition on the matrix element of the quasi-PDF in an off-shell quark state $|p,s\rangle$:

\begin{align}
&\tilde{Z}(z,p^R_z,a^{-1},\mu_R)\nonumber\\
&=\left.\frac{\textrm{Tr}[\slashed p \sum_s \langle p,s|O_{\gamma_t}(z)|p,s\rangle]}{\textrm{Tr}[\slashed p  \sum_s \langle p,s|O_{\gamma_t}(z)|p,s\rangle_{tree}]}\right|_{\tiny\begin{matrix}p^2=-\mu_R^2 \\ \!\!\!\!p_z=p^R_z\end{matrix}}
\end{align}
On the lattice, $\langle p,s|O_{\gamma_t}(z)|p,s\rangle$ is calculated from the amputated Green function of $O_{\gamma_t}$ with Euclidean external momentum.

The renormalization constant $\tilde{Z}(z,p^R_z,a^{-1},\mu_R)$ depends on the lattice spacing as well as the other two scales $p_z^R$ and $\mu_R$. It is used to renormalize the bare nucleon matrix element of the quasi-PDF $\tilde{h}(z,P_z,a^{-1}) = \frac{1}{2P^0} \langle P |O_{\gamma_t}(z)| P \rangle$
in coordinate space,

\begin{align} \label{eq:rimomh}
&\tilde{h}_R(z,P_z, p^R_z,\mu_R)\\
&=\left.\tilde{Z}^{-1}(z,p^R_z,a^{-1},\mu_R)\tilde{h}(z,P_z,a^{-1})\right|_{a\to0}\nonumber.
\end{align}
Note that the continuum limit of $\tilde{h}_R(z,P_z, p^R_z,\mu_R)$ is well defined and should be taken before the matching to the PDF.
Consequently, the RI/MOM quasi-PDF $\tilde{q}(x,P_z, p^R_z,\mu_R)$ is obtained through the Fourier transform of $\tilde{h}_R(z,P_z, p^R_z,\mu_R)$,

\begin{equation}
	\tilde{q}(x,P_z, p^R_z,\mu_R) = P_z\int {dz\over 2\pi}\ e^{ix P_zz}\tilde{h}_R(z,P_z, p^R_z,\mu_R)\,.
\end{equation}

$\tilde{h}_R(z,P_z, p^R_z,\mu_R)$ and $\tilde{q}(x,P_z, p^R_z,\mu_R)$ are independent of the UV regulator, so the matching between the quasi-PDF and $\overline{\text{MS}}$ PDF can be carried out in the continuum theory with dimensional regularization. In Refs.~\cite{Constantinou:2017sej,Alexandrou:2017huk,Alexandrou:2018pbm}, $\tilde{h}_R(z,P_z, p^R_z,\mu_R)$ was first converted to the $\overline{\text{MS}}$ scheme in coordinate space, and then Fourier transformed into momentum space to obtain the $\overline{\text{MS}}$ quasi-PDF, and finally matched to the $\overline{\text{MS}}$ PDF with the matching coefficient from Refs.~\cite{Izubuchi:2018srq}.

Instead of performing a two-step matching, we choose to directly match $\tilde{q}(x,P_z, p^R_z,\mu_R)$ to the $\overline{\text{MS}}$ PDF, and the matching coefficient for the $O_{\gamma_z}$ case has already been calculated at one-loop order in perturbation theory~\cite{Stewart:2017tvs}. Although in principle it is equivalent to the two-step procedure described above, the direct matching can possibly save us from additional systematic uncertainties when we implement them numerically on the lattice data. The efficiency of both procedures will be compared in the end to check consistency.

Following the framework described in Refs.~\cite{Stewart:2017tvs,Izubuchi:2018srq}, the matching between quasi-PDF $\tilde{q}(x,P_z, p^R_z,\mu_R)$ and the $\overline{\text{MS}}$ PDF $ q(y,\mu)$ at scale $\mu$ we obtained is,
\begin{align} \label{eq:momfact}
\tilde{q}(x,P_z, p^R_z,\mu_R)= &\int_{-1}^1 {dy\over |y|}\: C\left({x\over y},r,\frac{yP_z}{\mu},\frac{yP_z}{p_z^R}\right) \, q(y,\mu)\nonumber\\
&+\mathcal{O}\left({M^2\over P_z^2},{\Lambda_{\text{QCD}}^2\over P_z^2}\right),
\end{align}
where $r={\mu_R}^2/{p^R_z}^2$, and we have absorbed the antiquark distribution $\bar{q}(y,\mu)=-q(-y,\mu)$ into the region $-1<y<0$. 
The matching coefficient $C$ in this work is calculated at one-loop level and will be presented in a future publication~\cite{long}. To predict the PDF using the quasi-PDF obtained by lattice calculation, we invert the matching Eq.~\ref{eq:momfact} by changing the sign of $\alpha_s$ in $C$ to obtain the matching coefficient, which is to be convoluted with $\tilde{q}$ to obtain the prediction $q$. 

{\bf Lattice-QCD Calculation Setup:}
In this paper, we report the results of a lattice-QCD calculation using clover valence fermions on an ensemble of gauge configurations with lattice spacing $a=0.09$~fm, box size $L\approx 5.8$~fm and pion mass $M_\pi \approx 135$~MeV with $N_f=2+1+1$ (degenerate up/down, strange and charm) flavors of highly improved staggered quarks (HISQ)~\cite{Follana:2006rc} generated by MILC Collaboration~\cite{Bazavov:2012xda}. The gauge links are hypercubic (HYP)-smeared~\cite{Hasenfratz:2001hp} and then clover parameters are tuned to recover the lowest pion mass of the staggered quarks in the sea~\cite{Rajan:2017lxk,Bhattacharya:2015wna,Bhattacharya:2015esa,Bhattacharya:2013ehc}.
Only one step of HYP smearing is used to improve the discretization effects; too much smearing may alter the ultraviolet results for the PDF. 
We use Gaussian momentum smearing~\cite{Bali:2016lva} for the quark field 
\begin{align}
\psi(x) \rightarrow 
\psi(x) 
+ \alpha \sum_j U_j(x)e^{ik\hat{e}_j}\psi(x+\hat{e}_j),
\label{eq:moms}
\end{align}
where $k$ is the desired momentum,
$U_j(x)$ are the gauge links in the $j$ direction, 
and 
$\alpha$ is a tunable parameter as in traditional Gaussian smearing.
Such a momentum source is designed to align the overlap with nucleons of the desired boost momentum,
and we are able to reach higher boost momentum for the nucleon states than our previous work at physical pion mass~\cite{Lin:2017ani}.
In our previous exploratory study, although we varied our Gaussian smearing radius to better overlap with the largest momentum used in the calculation, the smearing of the field is still centered around zero in momentum space. 
With momentum smearing, the center of the smearing will be shifted to momentum $k$, which will immediately allow us to reach higher boost momenta with better signal-to-noise ratios in the matrix elements. 


To better control the systematics due to contamination by nearby excited states, 
we vary the Gaussian smearing parameter, $\alpha$, to better capture the excited state, leaving the signal for the ground-state nucleon cleaner. 
In addition, we use a simultaneous fit of the nucleon matrix element correlators,
using four source-sink nucleon separations, 0.72, 0.81, 0.90, 1.08~fm; 
the detailed procedure is described in Ref.~\cite{Bhattacharya:2013ehc} for the nucleon charges. 
We use multigrid algorithm~\cite{Babich:2010qb,Osborn:2010mb} in Chroma software package~\cite{Edwards:2004sx} to speed up the physical pion mass clover fermion inversion on the quark propagator.
We use multiple values of nucleon boost momenta, $\vec{P}=\{0,0, n \frac{2\pi}{L}\}$, with 
$n \in \{10,12,14\}$, which correspond to 2.2, 2.6 and 3.0~GeV nucleon momenta. 

\begin{figure}[htbp]
\includegraphics[width=.5\textwidth]{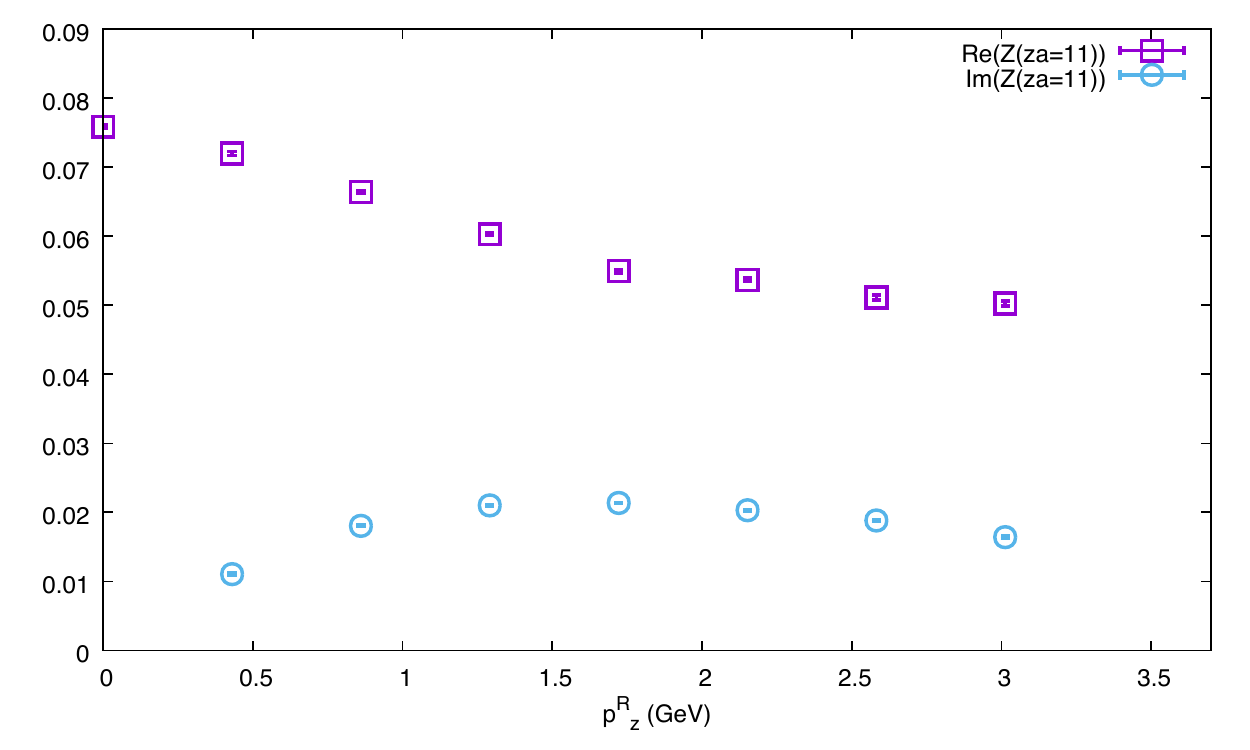}
\caption{The values of $Z(z)$ (the inverse of the renormalization constant) at $\mu_R=3.7$~GeV and $z=11 a \approx 1.0$~fm as a function of $p^R_z$. Note that $Z(z)$ becomes stable at large $p^R_z$.} \label{fig:pz_dependence}
\end{figure}

We apply RI/MOM scheme nonperturbative renormalization as described in our earlier work~\cite{Chen:2017mzz} but switch the $\Gamma$ to $\gamma_t$ to avoid operator mixing. We use momentum source for NPR propagators with quark momentum ranging in  $[0,14\times 2\pi/L]$. We study the momentum dependence using $\mu_R=2.3$ and 3.7~GeV, as well as the quark momentum dependence $p^R_z$. Fig.~\ref{fig:pz_dependence} shows the $p^R_z$ dependence at fixed $\mu_R=3.7$~GeV
where the $Z(z)$ is the inverse of renormalization constant. We find in the small-$p_z^R$ region there is a notable change in renormalization constants, while at large $p^R_z$, it seems reach a plateau and become stable. Similar behavior is also observed in $\mu_R=2.3$~GeV case. Therefore, we pick $p^R_z$=$10\times 2\pi/L$ as our central value for the renormalization constant.

{\bf Parton Distribution Function Results and Discussion:}
We use the ``derivative'' method proposed in our earlier work~\cite{Lin:2017ani} to improve the truncation error due to the Fourier transformation into $x$ space; that is, we take the 
derivative of the renormalized nucleon matrix elements $\partial_z\tilde{h}_R(z)$ whose Fourier transformation differs from the original in a known way:
\begin{align}
\label{eq:derivative}
\tilde{q}(x) = \int_{-z_\text{max}}^{+z_\text{max}}dz \frac{e^{i x P_z z}}{i x} \partial_z\tilde{h}_R(z)
\end{align}
$\partial_z\tilde{h}_R(z)$ is consistent with zero for $|z|>15a$ and we took $z_\text{max}=20a$ in this work.
The residual truncation systematics can be quantified by using a known global PDF input by checking how well it reproduces itself at lattice parameters, as outlined in Ref.~\cite{Lin:2017ani}.

We also investigate the excited-state contamination in the PDF. Excited-state contamination is notorious for contaminating the well-known nucleon axial charges in many past lattice-QCD calculations. As we increase the nucleon boost momentum, we anticipate that excited-state contamination worsens, since the states are relatively closer to each other; therefore, a careful study of the excited-state contamination is necessary for the LaMET (or quasi-/pseudo-PDF) approach. 
We use multiple analysis methods to remove excited-state systematics among 4 source-sink separations used in this work:
First, we use the ``two-sim'' analysis described in Ref.~\cite{Bhattacharya:2013ehc} to obtain the ground-state nucleon matrix elements using all 4 source-sink separations. A second extraction uses this method but only the largest two separations.
Then, we use the ``two-simRR''analysis (see Ref.~\cite{Bhattacharya:2013ehc} for details), which includes an additional matrix element related to excited states but with almost doubling the errors comparing with the ``two-sim'' analysis. Our final matrix element when combining all the three analyses is consistent with ``two-simRR'' analysis. 
For the rest of this paper, we will focus on the results using ``two-simRR''; it has a larger statistical error due to the inclusion of excited states, but the result will be more reliable.

\begin{figure}[htbp]
\includegraphics[width=.4\textwidth]{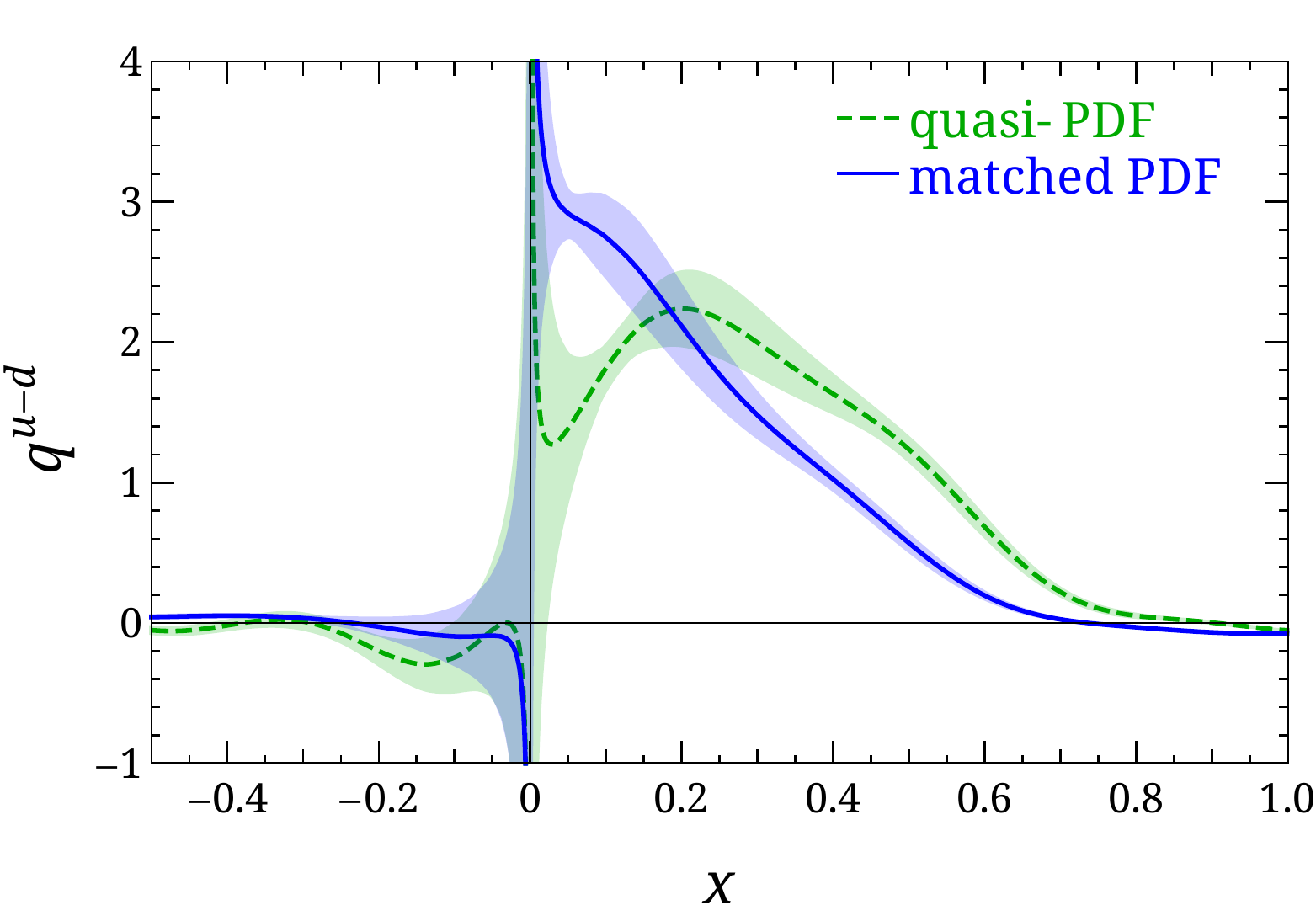}
\caption{The quasi-PDF and matched PDF with nucleon boost momentum 2.2~GeV. The parameters of the central value of matched PDF is $(\mu_R,p_z^R)=(3.7,2.2)$~GeV. The matching process lowers the quasi-PDF at large positive $x$ and enhances the small-$x$ region's quark asymmetry. It significantly changes the antiquark asymmetry at this nucleon momentum.} \label{fig:matchingPDF}
\end{figure}

\begin{figure}[htbp]
\includegraphics[width=.4\textwidth]{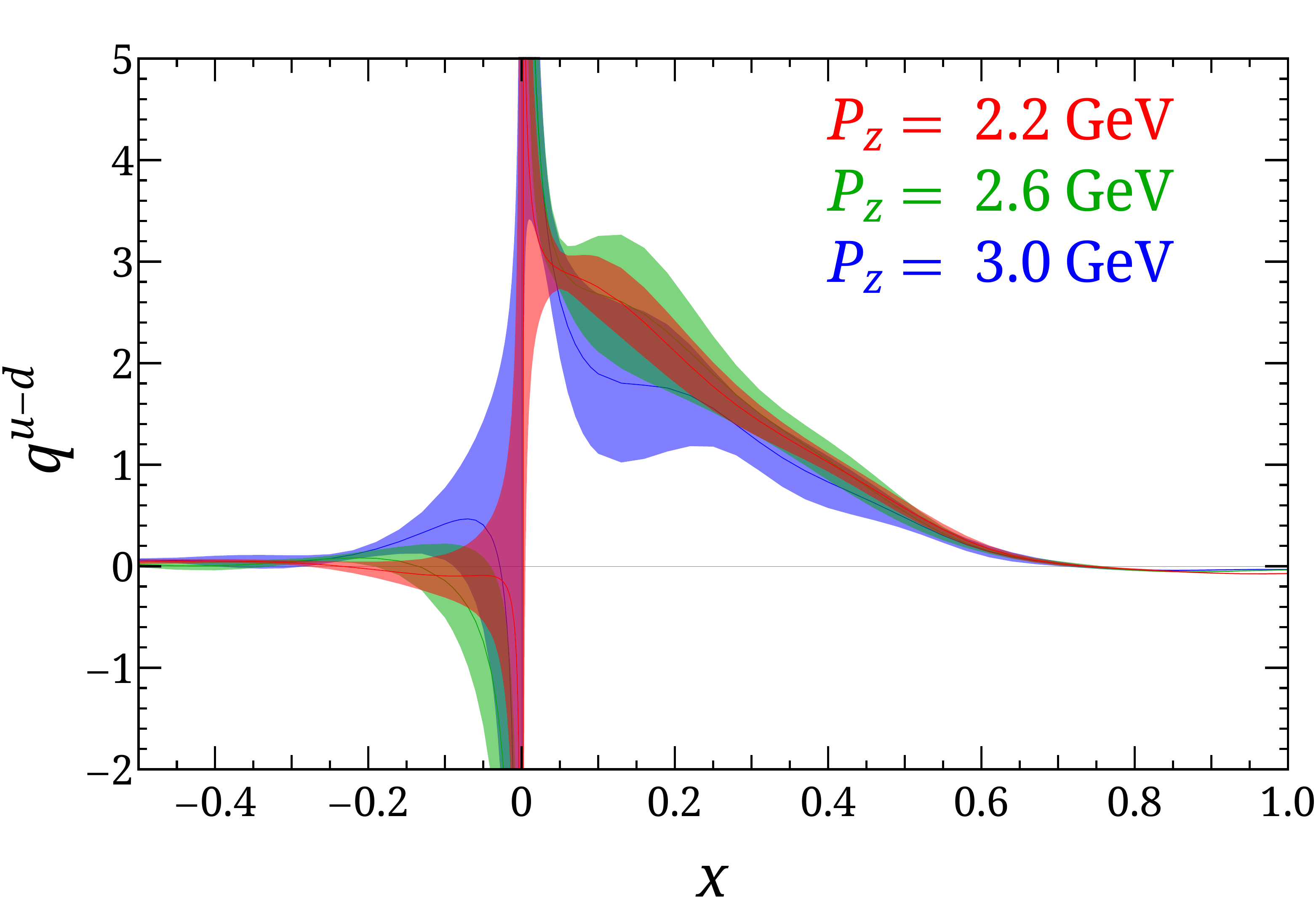}
\caption{Nucleon boost momentum dependence of the matched unpolarized isovector PDFs. The parameters of the central value of matched PDF is $(\mu_R,p_z^R)=(3.7,2.2)$~GeV. For quark asymmetry, the shape is consistent throughout most $x$ regions. However, in the antiquark region, there is a significant change in distribution as momentum increases.} \label{fig:matchingPDFallPz}
\end{figure}

Using the ``two-simRR'' renormalized PDF and applying the $\gamma_t$ matching, we show in Fig.~\ref{fig:matchingPDF}
 one of the three nucleon boost momenta, $P_z=2.2$~GeV, before and after applying the matching formula of Eq.~\ref{eq:momfact}. 
The matching raises the antiquark (i.e. negative-$x$ region) asymmetry for $x<-0.05$, and lowers the positive mid-$x$ to large $x$ quark distribution, compared with our exploratory study and heavier-pion PDF. 
After matching, we study the dependence on the nucleon boost momentum, shown in Fig.~\ref{fig:matchingPDFallPz}. Within the statistical errors, the distribution seems to converge across the three momenta. However, the central values shift noticeably from 2.2 to 3.0~GeV, moving the antiquark distribution toward the asymmetry measured in experiment: $\bar{d}(x) >\bar{u}(x)$.

\begin{figure}[htbp]
\includegraphics[width=.4\textwidth]{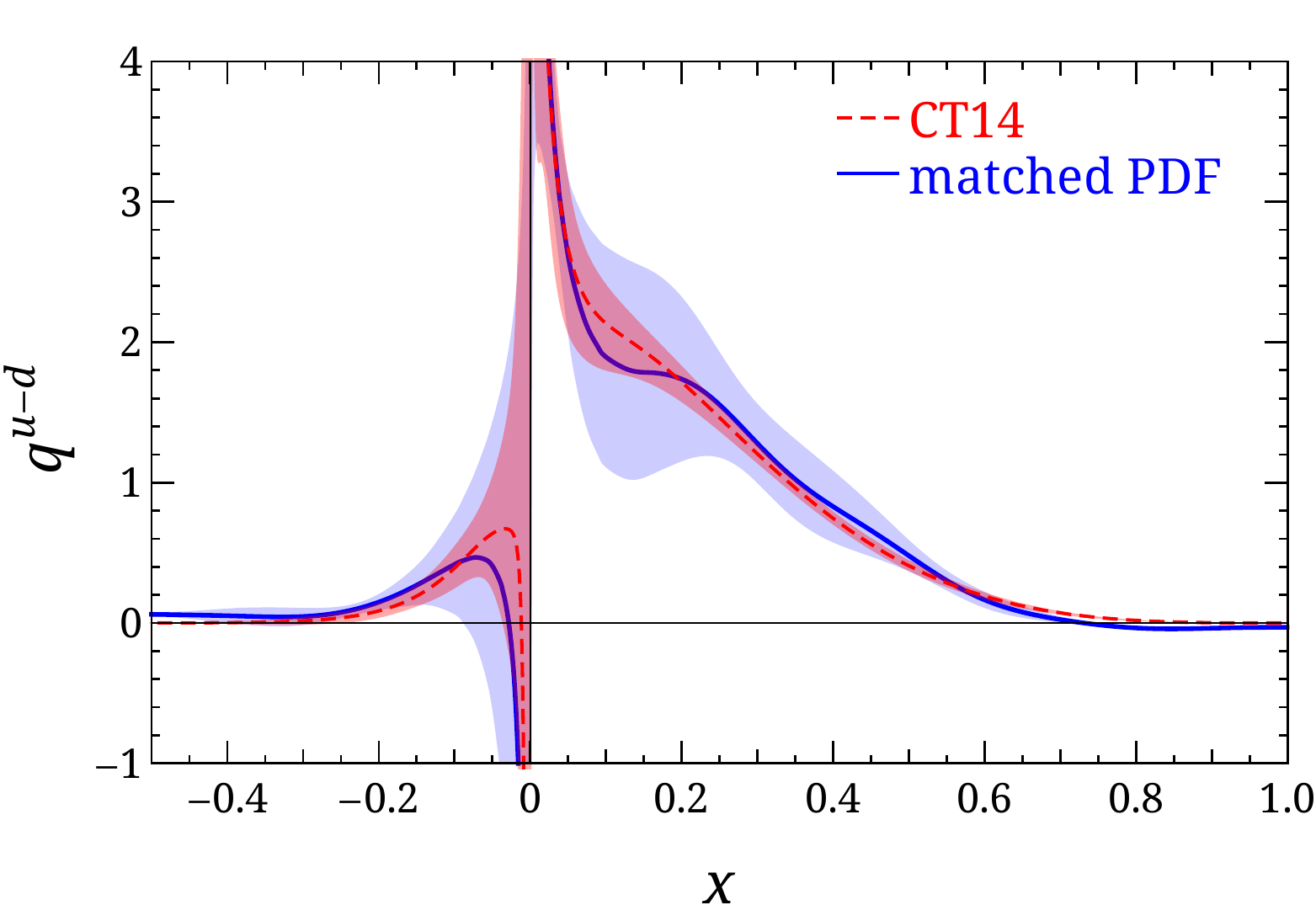}
\caption{Our final PDF renormalized at 3~GeV and compared with CT14~\cite{Dulat:2015mca} at $(\mu_R,p_z^R)=(3.7,2.2)$~GeV. It is consistent with NNPDF3.1 distribution~\cite{Ball:2017nwa} and CJ15~\cite{Accardi:2016qay}. 
Our results agree nicely with the global-analysis PDF.} \label{fig:finalPDF}
\end{figure}

The final result from this paper, shown in Fig.~\ref{fig:finalPDF}, significantly improves on our previous results at physical pion mass~\cite{Lin:2017ani}. We increase the nucleon momenta used in the calculation from 0.4, 0.8, 1.3~GeV to 2.2, 2.6 and 3.0~GeV. We use $O_{\gamma_t}$ operator here rather than the $O_{\gamma_z}$ used in Ref.~\cite{Chen:2017mzz}. (Though we showed in Ref.~\cite{Chen:2017mzz} that the final PDF is consistent when including the scalar matrix elements.) We extend our matrix element analysis to include an extra term due to the excited states. We increase at least a factor of 10 in statistics.  We also include the complete matching needed from the RI/MOM quasi-PDF to $\overline{\text{MS}}$ lightcone PDF.

While finalizing our analysis, another PDF result at physical pion mass was reported by ETMC~\cite{Alexandrou:2018pbm}. There are several advantages to our work: Larger $M_\pi L$ (roughly 4 vs 3) to minimize the finite-volume effects, larger nucleon boost momentum to suppress the power corrections, bigger $zP_z$ range to increase reliable $x$-regions, multistate fitting to remove excited-state contamination. It is notable that our results agree with the sea quark asymmetry seen experimentally and agree with zero in unphysical regions $|x|>1$. ETMC's results have oscillations that continue into the $|x|>1$ regions, which is possibly due to truncation of the Fourier transformation~\cite{Lin:2017ani}.

Other possible sources of systematic uncertainty include:
1) Higher-twist corrections: 
Taking $P_z=3.0$~GeV, we do not see significant difference in the $|x|>0.2$ regions, indicating that the higher-twist correction is well-controlled when $x$ is not too small.
2) Truncation effects:
At our largest nucleon boost momentum, our largest $z_\text{max}P_z$ is around 27.5. When we reconstruct a known PDF, there is significant difficulty in reproducing the small-$x$ regions $|x|<0.15$. Note that other lattice-PDF calculations use smaller $z_\text{max}P_z$, yielding smaller reliable regions in $x$-space. 
By extending to large displacement $z$ or large momentum $P_z$ on finer lattice spacing in future work, this can be improved straightforwardly.
3) $\mathcal{O}(\alpha_s^2)$ error in perturbation theory: we estimated the $\mathcal{O}(\alpha_s^2)$ error by applying matching to experimental fitted PDF to quasi-PDF using Eq.~\ref{eq:momfact} and then do the inverse matching back to PDF. We find that the error is about the same size as the statistical error.

We have presented new lattice-QCD results for the isovector unpolarized PDF (that is, the up-down quark asymmetry in the proton), which has much potential impact on current PDF estimates in the near future: 
1) The isovector PDF at large $x$ can be used as a constraint in global PDF analysis. The large-$x$ experimental data are often contaminated by nuclear effects, which are hard to cleanly remove. Many current PDF analyses (see references in Ref.~\cite{Lin:2017snn}) rely on extrapolation in these regions. There are ongoing LHCb measurements that can potentially improve and constrain the PDF in the large-$x$ region, but the precision of these data are not yet good enough to make a difference. A recent community whitepaper among lattice and global analysis practitioners~\cite{Lin:2017snn} predicted that a calculation of the large-$x$ isovector with 10\% final error can improve on the current PDF, especially in the antiquark regions where experimental inputs are even scarcer. Currently, we are able to reproduce the global PDF results; the next step will be to plan improved calculations with total uncertainty less than 10\%.
2) With the promising results shown here, we can proceed with similar analyses for the less known polarized PDFs, such as helicity and transversity (the longitudinal and transversely polarized PDFs), where the isovector PDFs needed to make impacts for global analysis are less demanding than the unpolarized ones.

{\bf Summary and Outlook:}
In this work, we report the state-of-the-art isovector unpolarized quark distribution using lattice QCD directly at physical pion mass. We use nucleon boosted momenta as large as 3~GeV with high-statistics analysis. We carefully study excited-state systematics whose error is reflected in our final distribution uncertainty. We renormalize our nucleon matrix element using the nonperturbative RI/MOM renormalization, and perform the LaMET one-loop finite-momentum matching and conversion to $\overline{\text{MS}}$-scheme to connect lattice quasi-distribution to lightcone distribution. We found our final distribution agree well with the global analysis distribution. We carefully examine all possible systematics which will give us better guideline to improve our future calculations and provide better precision distributions. 
Future direction will be investigating smaller lattice spacing ensembles for reaching even higher boosted momentum such that we can push toward smaller-$x$ region. 


\section*{Acknowledgments}
We thank the MILC Collaboration for sharing the lattices used to perform this study. The LQCD calculations were performed using the Chroma software suite~\cite{Edwards:2004sx}. 
This research used resources of the National Energy Research Scientific Computing Center, a DOE Office of Science User Facility supported by the Office of Science of the U.S. Department of Energy under Contract No. DE-AC02-05CH11231 
through ALCC and ERCAP;
facilities of the USQCD Collaboration, which are funded by the Office of Science of the U.S. Department of Energy,
and supported in part by Michigan State University through computational resources provided by the Institute for Cyber-Enabled Research. 
JWC is partly supported by the Ministry of Science and Technology, Taiwan, under Grant No. 105-2112-M-002-017-MY3 and the Kenda Foundation. LCJ is supported by the Department of Energy, Laboratory Directed Research and Development (LDRD) funding of BNL, under contract DE-EC0012704. YSL is supported by Science and Technology Commission of Shanghai Municipality (Grant No.16DZ2260200) and National Natural Science Foundation of China (Grant No.11655002). HL and YY are supported by the US National Science Foundation under grant PHY 1653405 ``CAREER: Constraining Parton Distribution Functions for New-Physics Searches''. JHZ is supported by the SFB/TRR-55 grant ``Hadron Physics from Lattice QCD'', and a grant from National Science Foundation of China (No.~11405104). YZ is supported by the U.S. Department of Energy, Office of Science, Office of Nuclear Physics, from DE-SC0011090 and within the framework of the TMD Topical Collaboration.

\bibliographystyle{apsrev}
\bibliography{ref}
\end{document}